\documentclass{IEEEtran}
\usepackage{times}  
\usepackage{graphicx} 
\usepackage{color}
\usepackage{pdfsync}
\usepackage{amsmath}
\usepackage{amssymb}
\usepackage{mathrsfs}

\usepackage{fix2col}
\usepackage{algpseudocode}
\usepackage{algorithm}
\newcommand{\eq}{\triangleq}

\newcommand{\field}[1]{\mathbb{#1}}
\newcommand{\R}{\field{R}}
\newcommand{\N}{\field{N}}

\newcommand{\B}{\field{B}}

\newcommand{\Prob}{\mathbf{Pr}}
\newcommand{\E}{\mathbf{E}}
\newcommand{\I}{\field{I}}

\newcommand{\Gauss}{\mathcal{N}}

\newcommand{\hfs}{\hfill\ensuremath{\square}}

\newcommand{\col}{\mathrm{col}}

\newtheorem{thm}{Theorem} 
\newtheorem{rem}{Remark}
 
\newtheorem{ex}{Example}
\newtheorem{prop}{Proposition}
\newtheorem{ass}{Assumption}

\title{Adaptive Controller Placement for Wireless Sensor-Actuator Networks with
  Erasure Channels}  
 
\author{Daniel~E.~Quevedo,
   Karl H.\ Johansson, Anders~Ahl\'en, and Isabel Jurado%
\thanks{%
    Daniel Quevedo is
  with the School of Electrical Engineering \&
  Computer Science, The University of Newcastle, NSW
  2308, Australia; dquevedo@ieee.org. Karl H.\ Johansson is with ACCESS Linnaeus Centre, School of Electrical
  Engineering, Royal Institute of 
  Technology, Stockholm, Sweden; kallej@ee.kth.se.
  Anders~Ahl\'en is with Signals and Systems, Uppsala University,  SE-751 21,
  Uppsala, Sweden; Anders.Ahlen@signal.uu.se. 
  Isabel Jurado is with the Departamento de Ingenier\'{\i}a de Sistemas y Autom\'atica,
  Escuela Superior de Ingenieros, Universidad de Sevilla, Spain. ijurado@cartuja.us.es}
\thanks{Research  supported  by Australian
  Research Council's 
  Discovery Project DP0988601, 
  by VINNOVA project WiComPI, project Dnr2009-02963, by MCyT (Grant
  DPI2010-19154), and by the European Commission FeedNetBack Project (Grant
  223866). A preliminary version
  of parts of this work was presented as\cite{queahl11b}.}}

\begin{document}

\maketitle

\begin{abstract}                         
Wireless sensor-actuator networks offer flexibility for control design. One
novel element which may arise in networks with multiple
nodes is that the
role of some  nodes does not need to be fixed. In particular,  there is no 
need to pre-allocate  which
nodes assume 
controller functions and which ones merely relay data. We present a flexible 
architecture for networked control using multiple nodes connected in series
over analog erasure channels without acknowledgments. The control
architecture proposed adapts to changes in 
network conditions, by allowing the role played by individual nodes to depend
upon transmission outcomes. We adopt stochastic models for transmission
outcomes and characterize the distribution of controller location and
the covariance of system states. Simulation results illustrate that the proposed
architecture has the potential to give better  performance than
limiting control calculations to be carried out at a fixed node.
\end{abstract}

\section{Introduction}
\label{sec:introduction}
In a Networked Control System (NCS), sensor, controller and actuator links are
not transparent, but are affected by
bit-rate limitations, packet dropouts and/or delays. This  leads to
performance degradation and
makes the design of NCSs often a challenging 
task \cite{antbai07,chejoh11}.   An interesting aspect is that, when
compared to traditional 
hard-wired control loops, wireless NCSs offer  
 architectural flexibility and
additional degrees of 
freedom.
Whilst sensor and actuator functionalities will generally be pre-allocated, often there is no need to pre-assign in a static fashion which nodes carry out 
control calculations, and which nodes merely relay data. Intuitively,  the roles
of individual nodes should depend on the information available at each time instant.
 In the present work, we
examine this  question for the case of NCSs with random packet dropouts.  Our
motivating application is real-time control in the process industry. Several new
standards have recently been introduced for multi-hop wireless sensor and
actuator networks, e.g., WirelessHART, ISA-100, and this paper proposes a new
adaptive controller placement suitable to be implemented over  these
standards. Note that the plant time constants in process industry are often of the
order of seconds or minutes (or even higher), so we make the reasonable
assumption that network-induced delays can be neglected. 
 As background to our current work,   \cite{gooque08} studies performance of three static NCS
architectures by adopting an additive signal-to-noise
ratio constrained channel model. The  results in \cite{gooque08} suggest that, in the
absence of coding, placing the controller at the actuator node will give better
performance than placing it at the sensor node.   The work \cite{robkum08} examines
NCSs with stochastic packet dropouts using optimal control
techniques. \emph{Inter-alia}, the work shows that    
  optimal   performance can be achieved if all nodes aggregate
  their entire history of received data and relay it
  to the controller-actuator.  Depending upon the
  information available at each node, various optimal control 
  problems can be analyzed. More recently,
  \cite{pajsun11} investigates a distributed control strategy wherein the
  network 
  itself acts as a controller for a MIMO plant. All nodes (including the
  actuator nodes) perform linear
  combinations of internal state variables of neighboring nodes. In the case of
  analog erasure channels with i.i.d.\ 
  dropouts (without acknowledgments), in \cite{pajsun11} the resulting NCS is
  then cast,  
  analyzed and  designed as
  a jump-linear system. 

\par The 
  present work studies a single-loop
  NCS topology which uses a series 
  connection of analog erasure channels.  We focus on situations where the wireless nodes 
 have only limited energy and processing power,  precluding  long
data packets. Further,  the nodes do not provide local transmission
acknowledgments. Feedback (or acknowledgment) is only provided by the actuator node, which
broadcasts the applied plant input value over parallel unreliable links to the
intermediate nodes. This strategy is plausible as the actuator node is powered,
which is a reasonable assumption for most actuators in process industry.   Due
to random dropouts,   the actuator node does not
  have full 
  information on plant outputs. This  constitutes one of the major
  difficulties when implementing a controller in such a NCS. We address this
  issue by using an estimation and control structure which is distributed
  across the network. Instead of tackling optimal control formulations (which depend
upon network parameters and may therefore be difficult to implement in practice), we  adopt
a so-called emulation-based approach, where the controller has been
pre-designed; see, e.g., \cite{nestee04a,donhee11a,anthes13}.  To be more specific, we assume
that the control policy consists of a pre-designed state feedback-gain combined
with a state observer, which,
in the absence of network effects, would lead to the desired performance. Within
this context, we present a flexible NCS architecture where the role played by
individual nodes depends upon transmission outcomes. While all nodes calculate
local state estimates at all times, with 
the  algorithm proposed, transmission outcomes determine, 
at each instant, whether the control input will be calculated at the actuator
node, at the sensor node or at one of the intermediate nodes. It turns out that, if individual  dropout
processes are i.i.d., then the controller location has a stationary distribution, which can be
easily characterized. To analyze the performance of the dynamic NCS 
architecture in the presence of correlated dropouts, we derive a
jump-linear system model and adopt the network model recently introduced
in \cite{queahl13a}. This model encompasses temporal and spatial correlations  of
packet dropouts and is therefore of more practical importance than more
traditional i.i.d.\ models. The present paper goes beyond  our recent conference
contribution \cite{quejoh12a}, by presenting  a closed loop
model and considering networks
with correlated dropouts. Our approach complements \cite{quejoh12a} by
focusing on a nominal linear 
estimator, as opposed to a time-varying Kalman filter. This opens the
possibility to analyze NCS performance with correlated dropouts
using techniques from jump-linear systems.

\par This paper is organized as
follows. Section~\ref{sec:single-loop-control} describes the NCS topology of
interest. In Section~\ref{sec:ad-hoc-algorithm} we present the dynamic NCS
architecture. For the case of independent dropouts, Section~\ref{sec:analysis-}
provides the distribution of the controller
location. Section~\ref{sec:further-analysis} derives an overall closed loop
model. Performance issues are studied in
Section~\ref{sec:wnat-next}. Section~\ref{sec:simulation-study} documents
simulation studies. Section~\ref{sec:conclusions} draws conclusions.

\paragraph*{Notation}
\label{sec:notation}
We write 
$\N_0$ for $\{0, 1, 2, \ldots\}$; 
$\R$ are the real numbers, whereas $\R_{\geq 0}\eq [0,\infty)$. 
The $\ell$-th unit row-vector in Euclidean space is denoted $e_\ell$, for
example,  $e_2=
[\begin{matrix}
  0 & 1 & 0 & \dots & 0
\end{matrix}]$; $I_n$ is the
$n\times n$ unit matrix, $0_n\eq 0\cdot I_n$; $\otimes$
refers to the Kronecker product. For any set of column vectors,
$\{u_1,\dots,u_n\}$, $\col(u_1,\dots,u_n)=[u_1^T,\dots,u_n^T]^T$.
We  adopt the convention $\sum_{j=1}^{0}a_j =0$, for all $a_0,a_1\in\R$.  
 A real random variable $\mu$, which is zero-mean Gaussian with  covariance
  $\Gamma$ is denoted by $\mu \sim\Gauss(0, \Gamma)$.

\begin{figure}[t]
  \centering
    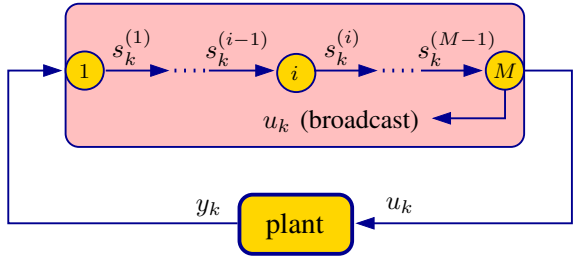
    \caption{Single-loop control over a wireless sensor-actuator network:
      forward- and feedback-links are unreliable.}
    \label{fig:single_loop}
\end{figure}

\section{Wireless Sensor-Actuator Network Setup}
\label{sec:single-loop-control}
We consider MIMO LTI plant models of the form
\begin{equation}
  \label{eq:1}
  \begin{split}
    x_{k+1}&=Ax_k +Bu_k+w_k\\
    y_k&=Cx_k+v_k,\quad  k\in\N_0
  \end{split}
\end{equation}
where $x_0 \sim\Gauss(\bar{x}_0,P_0)$, $P_0\succ 0$. In~\eqref{eq:1}, $u_k\in\R^m$ is
the plant
input, $x_k\in\R^n$ is the state,  $y_k\in\R^p$ 
is the output, and $w_k\sim\Gauss(0,Q)$, $Q\succ 0$ and
$v_k\sim\Gauss (0,R)$, $R\succ 0$,  are driving noise and measurement noise,
respectively.
As foreshadowed in the introduction, we focus on a situation where  suitable
  feedback and estimator gains $L\in\R^{m\times n}$ and $K\in\R^{n\times p}$
  have been pre-designed for a situation where the controller $(K,L)$ has perfect access to plant
  outputs and inputs. Consequently, we assume that if the control inputs 
 \begin{equation}
   \label{eq:11}
    u_k=L\hat{x}_k^{\text{nom}},\quad k\in\N_0, \quad\text{with}
 \end{equation}
\begin{equation}
  \label{eq:29}
  \hat{x}_{k}^{\text{nom}}= A\hat{x}_{k-1}^{\text{nom}} +
    B {u}_{k-1}+K\big(y_k - C ( A\hat{x}_{k-1}^{\text{nom}} +
    B {u}_{k-1})\big),
\end{equation}
where $\hat{x}_{k}^{\text{nom}}$ denotes an estimate of the state $x_k$,
were implemented at the plant, then  satisfactory performance would be
attained. The main theme of the present work is to investigate how to implement
the above nominal controller, when using a wireless sensor-actuator network with
dropouts.

\par
\begin{figure}[t]
  \centering
    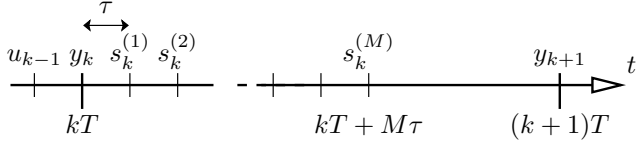
    \caption{TDMA Transmission Schedule. The
       plant input $u_k$ is applied at time $kT+M\tau$ and broadcast at time $(k+1)T-\tau$.}
    \label{fig:timeline}
\end{figure}

 The sensor node measures the plant output $y_k$, whereas the actuator node
manipulates the plant input $u_k$. The loop is closed over a wireless network,
characterised via a (directed) line-graph having $M$ nodes, see
Fig.~\ref{fig:single_loop}. 
Transmissions
are in
sequential Round-Robin fashion $\{1,2,\dots, M, 1, 2, \dots\}$ as depicted in
Fig.~\ref{fig:timeline}. More precisely, the packet $s_k^{(i)}$ is transmitted
from node ${i}$ to node 
${i+1}$  at times $kT+ i\tau$, where $T$ is the sampling period of~\eqref{eq:1}
and $\tau \ll T/(M+1)$  refers to the times between transmissions of
      packets $s_k^{(i)}$. The   input $u_k$ is applied at time
      $kT+(M+1)\tau$. We thus assume that
in-network processing is 
much faster than the plant 
dynamics~\eqref{eq:1} and neglect delays introduced by  the network.

\par The network
introduces stochastic packet dropouts. To study the situation, we adopt an analog erasure
channel model and introduce the binary success  processes
$$\gamma_k^{(i)}\in\{0,1\}, \quad k\in\N_0,\; i\in\{1,2,\dots,M-1\},$$
where $\gamma_k^{(i)}=1$ indicates that
 transmission of the packet $s_k^{(i)}$ from node $i$ to  node
${i+1}$  at time $kT+ i\tau$, is
successful, i.e., error-free;  $\gamma_k^{(i)}=0$ refers to a packet-dropout. 
Throughout this work,
we assume that transmission outcomes are known at the corresponding receiver
sides and that the sensor node $i=1$
has 
direct access to plant output 
measurements. For notational convenience, we write $\gamma_k^{(0)}=1$, for all
$k\in \N_0$. To save energy,  the
wireless nodes $i\in\{1,2,\dots, M-1\}$ do not provide acknowledgments of receipt of
the packets. The actuator
node $M$ provides a feedback
mechanism: At time $(k+1)T-\tau$, it broadcasts the control value $u_k$ to 
nodes $i\in\{1,\dots,M-1\}$, see Fig.~\ref{fig:single_loop}. Due to channel
fading, the
\emph{feedback links} between actuator and sensors are also
affected by dropouts. We denote the associated
success processes via
$$\delta_k^{(i)}\in\{0,1\},\quad k\in\N_0,\;i\in\{1,2,\dots,M-1\}.$$ More
precisely, if $u_{k}$ is successfully  
received at node $i$, then we set $\delta_{k}^{(i)}=1$; see also
\cite{imeyuk06} for studies on the importance of
acknowledgments in closed loop control. We assume  that the actuator node has 
perfect knowledge 
of plant inputs, and
thus, write $\delta_k^{(M)}=1$, $\forall k\in \N_0$.  Since the actuator node
will, in general, have less stringent energy
constraints than the other nodes, we focus our attention on situations where the
feedback links are more reliable than the forward links moving data from the
sensor to the actuator.

\par Due to packet dropouts, plant output measurements are not always
available at the actuator node. On the other hand, the sensor node will, in general, not have
perfect information of previous plant inputs. This makes the implementation  of
the nominal controller~\eqref{eq:11}--(\ref{eq:29}) a challenging task. 
In the sequel, we will present an adaptive controller
placement algorithm
where the computations leading to the plant inputs are distributed across the
network.  We foresee that our
approach will lead to a dynamic assignment of the role played by the individual network
nodes. As will become apparent in the sequel, which tasks are carried out by individual nodes at each time instant, will
depend upon transmission outcomes.


\section{Flexible Controller Placement}
\label{sec:ad-hoc-algorithm}
To keep  communication overheads low, the packets
transmitted by each node $i$ have only two fields, namely,
output measurements  and  tentative plant inputs  (if available):
\begin{equation}
  \label{eq:30}
 s_k^{(i)}= \big(y_k,    u_k^{(i)}\big).
\end{equation}
Plant outputs are transmitted in order to pass on information on the plant
state to the  nodes $\{i+1,i+2,\dots,M\}$, see Fig.~\ref{fig:single_loop}. On
the other hand, 
$u_k^{(i)}$ in~(\ref{eq:30}) is the plant input which is applied at the plant
provided the packet $s_k^{(i)}$ is delivered at the actuator node. If  $s_k^{(i)}$ is
lost, then following the algorithm described below,  the plant input will be provided by one of
the later nodes $\ell>i$, which thereby takes on the controller
role at time $k$.   In the sequel, we will refer to the node which
calculates the plant input at time $k$ as $c_k\in\{1,2,\dots,M\}$:
\begin{equation} 
  \label{eq:12}
  u_k=L\hat{x}_k^{(c_k)},\quad k\in\N_0,
\end{equation}
where $\hat{x}_k^{(c_k)}$ is the local plant state estimate computed at  node
$c_k$.
Intuitively, good control
performance will be achieved if the estimate used in~(\ref{eq:12}) is
accurate. Clearly, due to the multi-hop nature of the network,
nodes which are closer to the sensor will have access to more output
measurements, see Fig.~\ref{fig:single_loop}. On the
other hand, one can expect that nodes which are physically located closer to the
actuator node will on average receive more plant input
acknowledgments, thus, have better knowledge of plant inputs.

\par
While only the node $c_k$ will provide  $u_k$, in our
formulation all nodes
compute local state estimates, $\hat{x}_k^{(i)}$, by using the data received
from the actuator node and the preceding node. This serves as safeguard for instances when the loop is
broken due to dropouts. 
Motivated by the fact that often feedback links from the actuator to the
intermediate sensors are 
``quite reliable'', we adopt the following simple procedure: If plant
output measurements are available at node $i$, then state estimators are of the form~(\ref{eq:29}); at instances when the plant input
is not available, an open loop estimate is used, thus:
\begin{equation}
  \label{eq:15}
    \hat{x}_{k}^{(i)} = A\hat{x}_{k-1}^{(i)} +
    B\hat{u}_{k-1}^{(i)}+K_k^{(i)}\big(y_k - C ( A\hat{x}_{k-1}^{(i)} +
    B\hat{u}_{k-1}^{(i)})\big),
\end{equation}
where
\begin{equation}
  \label{eq:27}
    K_k^{(i)}\eq\Gamma_k^{(i)}K,\quad \text{and} \quad
   \Gamma_k^{(i)}\eq\prod_{j\in\{0,1,\dots,i-1\}} \gamma_k^{(j)}
 \end{equation}
 is equal to
  1 if and only if $y_k$
  is available at node $i$ at time $kT+(i-1)\tau$. In~\eqref{eq:15},
  $\hat{u}_{k-1}^{(i)}$ is a local plant input estimate. In particular, if
$\delta_{k-1}^{(i)}=1$, then  $\hat{u}_{k-1}^{(i)}=u_{k-1}$.
At instances where $\delta_{k-1}^{(i)}=0$,~node $i$ uses
 ${u}_{k-1}^{(i)}$, the tentative plant input value
  transmitted  in the second field of the previous packet $s_{k-1}^{(i)}$ (if
  non-empty), or otherwise sets $\hat{u}_{k-1}^{(i)}=L\hat{x}_{k-1}^{(i)}$. More
  details on the estimator are given in Section   \ref{sec:further-analysis}. 

\begin{rem}
Of course,
  the above  transmission and control strategy will in general not be
  optimal. In particular, nodes do
  not transmit local state estimates and the control law does not depend upon
  network parameters, e.g., dropout probabilities; cf., \cite{chisch11}. The
  aim of the present work is to 
  develop a simple and practical method, which uses an
  existing control and estimation policy  for implementation over an unreliable
  network and only requires little communication. \hfs
\end{rem}

\begin{rem}
  In \cite{quejoh12a}, instead of using~(\ref{eq:27}), the gains
  $K_k^{(i)}$ were taken  
as the Kalman filter gains for a system with intermittent
observations,
see, e.g., \cite{huadey07,queahl10,queahl12}.
Our subsequent analysis up to~(\ref{eq:26}) can be applied to this
structure as well. However, the jump-linear model derived in
Section~\ref{sec:further-analysis} requires a jump-linear estimation model, such
as~(\ref{eq:27}).\hfs
\end{rem}

\def\baselinestretch{1.2} 
\begin{algorithm}[h!]
  \caption{Adaptive Controller Placement}
  \label{alg1}
  \begin{algorithmic}[1]
\State  $k\gets 0$, $\hat{x}_0^{(i)}\gets 0$, $j\gets 0$ 
\While{$t\geq 0$} \Comment{$t\in\R_{\geq 0}$ is actual time}
\While{$t\leq kT+(i-1)\tau$} \Comment{wait-loop}
   \State $j\gets j+1$  
\EndWhile     
\If{$\gamma_k^{(i-1)}=0$} \Comment{$s_{k}^{(i-1)}$ is dropped}
   \State  $\hat{u}^{(i)}_{k} \gets L \hat{x}_{k}^{(i)}$ 
    \If{$\delta_{k-1}^{(i)}=1$} 
        \State \label{alg:c1} $s_{k}^{(i)}\gets \big(\emptyset,\hat{u}^{(i)}_{k}\big)$ 
        \Comment{a tentative  input}
        \Else \State \label{alg:empty}  $s_{k}^{(i)}\gets \big(\emptyset,\emptyset\big)$
    \EndIf
\EndIf

\If{$\gamma_k^{(i-1)}=1$} \Comment{$s_k^{(i-1)}$ is received}
  \State $(y,u)\gets s_k^{(i-1)}$
  \If{$y \not = \emptyset$} \Comment{$y_k$ is available}
  \State $\hat{x}_{k}^{(i)} \gets \hat{x}_{k}^{(i)} +K\big(y -
  C \hat{x}_{k}^{(i)}\big)$
\EndIf
 
\If{$u \not = \emptyset$}
\State ${u}^{(i)}_{k} =u$
\Else \State  ${u}^{(i)}_{k} \gets L \hat{x}_{k}^{(i)}$ 
\EndIf

\If{$u=\emptyset \land \delta_{k-1}^{(i)}=1$}
\State \label{alg:c2} $s_{k}^{(i)}\gets \big(y,{u}^{(i)}_{k}  \big)$
  \Comment{a tentative input}
\Else 
\State \label{alg:fwd} $s_{k}^{(i)}\gets (y,u)$ \Comment{$s_{k}^{(i-1)}$ is forwarded}

\EndIf
  \EndIf
  \While{$t<kT+i\tau$}  \Comment{wait-loop}
  \State $j\gets j+1$ 
  \EndWhile
  \State \textbf{transmit}   $s_{k}^{(i)}$
  \While{$t\leq (k+1)T-\tau$}  \Comment{wait-loop}
  \State $j\gets j+1$ 
  \EndWhile
  \If{$\delta^{(i)}_{k}=1$}  \Comment{ $u_{k}$ is available}  
    \State $\hat{x}_{k+1}^{(i)} \gets A\hat{x}_{k}^{(i)}+B u_{k}$
    \Else \Comment{the value in $s_{k}^{(i)}$ is used}
    \State $\hat{x}_{k+1}^{(i)} \gets A\hat{x}_{k}^{(i)}+B{u}^{(i)}_{k}$
    \EndIf
    
\State  $ k \gets k+1$
\EndWhile
  \end{algorithmic}
\end{algorithm}
\def\baselinestretch{1}

\par Algorithm~\ref{alg1}, run at every node $i\in\{1,2,\dots,M\}$, 
embodies the adaptive   controller allocation
method described in the preceding section. Which calculations are carried out at each node, depends upon
transmission 
outcomes involving the current node (see lines 6, 8, 14, 24 and 37) and also
transmission outcomes at previous nodes (see lines 16, 19 and 24).   In
particular,   node $i$ only calculates a 
tentative plant input when no tentative plant input is received from node $i-1$
and node 
$i$ has successfully received $u_{k-1}$ (see lines~\ref{alg:c1} and~\ref{alg:c2}). Therefore, 
preference is given  to  relay incoming tentative plant 
input values.
 The reason for adopting this decision procedure lies in that we assume that data sent from  the actuator node to intermediate nodes is often
available, whereas transmissions of packets $s_k^{(i)}$ are less
reliable. Thus, nodes closer to 
the sensor node can be expected to have better state 
estimates than nodes located  further down the line. 
 In particular, the sensor node $i=1$ uses as input 
\begin{equation}
  \label{eq:13}
  s_k^{(0)} =(y_k,\emptyset),\quad\gamma_k^{(0)}=1.
\end{equation}
If $\delta_{k-1}^{(1)}=1$, then 
the sensor node calculates a
tentative control value and 
transmits $s_k^{(1)}=(y_k,L\hat{x}_k^{(1)})$ to node
$2$. Subsequent nodes  relay this packet towards the actuator node. If the
packet is dropped along the way,  then the  next node  $i$ where $\delta_{k-1}^{(i)}=1$,
calculates a tentative 
control   $u_k^{(i)}=L\hat{x}_k^{(i)}$ and transmits
$s_k^{(i)}=(\emptyset,u_k^{(i)})$ to node $i+1$, etc.
On the other hand,  if  $\delta_{k-1}^{(1)}=0$, then
$s_k^{(0)}$ is relayed to subsequent nodes
until 
arriving at some node $i$ where $u_{k-1}$ was successfully received. Control calculations 
are then carried out and the packet $s_k^{(i)}$ obtained is
relayed towards the actuator node, etc. 
The actuator node implements  $u_k=u_k^{(M)}$, the value
contained in the 
second field of $s_k^{(M)}$.

\begin{rem}
 An advantage of allowing the control
calculations to be located  
arbitrarily and in a time-varying fashion, is that it makes more difficult for
someone to attack the NCS. The
development of  secure control strategies based on Algorithm 1 presented
remains a topic of future research.\hfs
\end{rem}

\section{Dynamic Controller Location}
\label{sec:analysis-}
With Algorithm~\ref{alg1}, which of the nodes calculates the plant input
$u_k$, depends upon the transmission
outcomes. 
For further reference, we shall denote  the set of nodes which calculate a
tentative control input (see lines~\ref{alg:c1} and ~\ref{alg:c2}  of Algorithm~\ref{alg1}) via
$\mathcal{C}_k\subset\{1,2,\dots,M\}.$ 
%
It is convenient to introduce the  processes
$\big\{\mu_k^{(i)}\big\}$, and $\big\{c_k^{(i)}\big\}$, where  
$i\in \{0,1,\dots,M\}$ and  
\begin{equation}
  \label{eq:9}
  \begin{split}
    \mu_k^{(i)} &\eq
    \begin{cases}
      0&\text{if the second field of $s_k^{(i)}$ is empty,}\\
      1&\text{otherwise,}
    \end{cases}\\
    c_k^{(i)}&\eq  \mu_k^{(i)}\max(\mathcal{C}_k \cap \{1,2,\dots, i\}).
  \end{split}
\end{equation}
Note that $\mu_k^{(1)}=\delta_{k-1}^{(1)},\forall k\in\N$. If $c_k^{(i)}>0$, then $c_k^{(i)}$ denotes the node where the second
field of $s_k^{(i)}$ was calculated.
 It is easy to see that, with the algorithm proposed and since the packets
$s_k^{(i)}$ are communicated  sequentially, see Fig.~\ref{fig:timeline}, we have
$c_k^{(1)}=\delta_{k-1}^{(1)}$, for all $k\in\N_0$, whereas 
\begin{equation}
  \label{eq:36}
  \begin{split}
    c_k^{(i)}=
    \begin{cases}
      i \delta_{k-1}^{(i)}&\text{if $c_k^{(i-1)}=0 \lor \gamma_k^{(i-1)}=0$,}\\
      c_k^{(i-1)}&\text{if $c_k^{(i-1)}>0 \land \gamma_k^{(i-1)}=1$,}
    \end{cases}
  \end{split}
\end{equation}
for $i\in\{2,\dots,M\}$, $k\in\N_0$. The  ``controller node at time $k$'', i.e., the node
where $u_k$ was calculated is given by
\begin{equation} 
  \label{eq:4}
  c_k \eq c_k^{(M)}=\max (\mathcal{C}_k),\quad \forall k\in\N_0,
\end{equation}
see~(\ref{eq:12}). 
To derive our results,  we introduce the aggregated transmission
outcome process $\{\beta_k\}$, $k\in\N_0$, where
\begin{equation}
  \label{eq:32}
  \beta_k\eq \sum_{i=1}^{M-1} \big( 2^{M-1}
  \gamma_{k+1}^{(i)}+\delta_k^{(i)}   \big)
2^{i-1},\quad k\in\N_0.
\end{equation}
Note that $\beta_{k-1}\in\I\eq\{0,1,\dots,2^{2M-2}-1\}$ collects the outcomes
of all transmissions which occur
 during the time-interval $[kT-\tau,kT+M\tau]$, see
 Fig.~\ref{fig:timeline}. Thus, $\beta_{k-1}$ determines $\mathcal{C}_k$ and
 $c_k$, %
i.e.,  the controller location will dynamically adapt to the
network conditions. To further 
elucidate the situation, in the sequel we will  regard $\{\beta_k\}$, $k\in\N_0$ as a stochastic process. We will first
assume that the transmission  processes are Bernoulli
distributed.  
\begin{ass}
\label{ass:processes}
The link transmission processes are independent and
identically 
distributed (i.i.d.) with a common success
  probability $p\in[0,1]$: 
  \begin{equation}
    \label{eq:7}
    \Prob\big\{\gamma^{(i)}_k=1\big\} = p, \quad \forall i\in\{1,2,\dots,M-1\}.
  \end{equation}
 The feedback link success processes are i.i.d., with 
  \begin{equation}
    \label{eq:5}
    \Prob\big\{\delta^{(i)}_k=1\big\} = q_i, \quad \forall i\in\{1,2,\dots,M-1\},
  \end{equation}
for given success probabilities $q_1,q_2,\dots,q_{M-1}\in[0,1]$. \hfs 
\end{ass} 
Note that while the above assumption imposes that
transmission processes are i.i.d., it does  take into account the
fact that radio connectivity from the 
actuator node to the other nodes will be distance dependent. It also does
not impose that the processes $\big\{\mu_k^{(i)}\big\}$, ${k\in\N_0}$ for different
nodes $i$ are independent. However, the assumption made does imply
stationarity, as apparent  from Proposition~\ref{lem:distr} given below. 

\begin{prop}
\label{lem:distr}
Suppose that Assumption~\ref{ass:processes} holds. Then
\begin{align}
  \label{eq:18}
  \Prob\big\{ \mu_k^{(i)}=1\big\} &= q_i +\sum_{j=1}^{i-1}p^jq_{i-j}
  \prod_{\ell=0}^{j-1} (1-q_{i-\ell})\\
  \label{eq:16}
  \Prob\{  i\in\mathcal{C}_k \} &=
  q_i\big( 1-  p\Prob\big\{\mu_k^{(i-1)}=1\big\}\big)\\
  \label{eq:21}
  \Prob\{c_k=i\}&=p^{M-i} \Prob \{i\in\mathcal{C}_k\},
\end{align}
for all $k\in\N_0$ and $i\in\{1,2,\dots,M\}$, and where  $q_M=1$. 
 \end{prop}

The above result shows how the distributions of 
$\mu_k^{(i)}$,   $\mathcal{C}_k$, and of $c_k$  depend upon the communication
success probabilities; i.e., the 
distribution of $\beta_k$, here modeled as i.i.d.

\begin{ex}
  Suppose that Assumption~\ref{ass:processes} holds and that the feedback links
  are always available,  that is, 
  $q_i=1$, for all $i\in\{1,\dots,M\}$.
 Expression~(\ref{eq:18}) then provides
  that $\Prob\{ \mu_k^{(i)}=1\} =1$, for all $i\in \{1,\dots,M\}$. Since, by~(\ref{eq:13}),
  $\Prob\{ \mu_k^{(0)}=1\} =0$, Proposition~\ref{lem:distr} gives that
  \begin{equation*}
    \Prob\big\{ 
  i\in\mathcal{C}_k \big\} =
  \begin{cases}
    1&\text{if $i=1$}\\
    1- p &\text{if $i\in \{2,\dots,M\}$}
  \end{cases}
  \end{equation*}
  and the
  controller location sequence has the following geometric-like distribution
  \begin{equation}
\label{eq:19}
    \Prob\{c_k=i\}=
    \begin{cases}
      p^{M-1} &\text{if $i=1$}\\
      (1-p)p^{M-i}&\text{if $i\in\{2,3,\dots,M\}$}.
    \end{cases}
  \end{equation}
On the other hand, if the actuator does not broadcast the plant input values at
all ($q_i=0$, $\forall i\not = M$), then, $\forall k\in\N_0$  
\begin{gather*}
  \Prob\{ \mu_k^{(i)}=1\} =0,\quad \forall i\in \{1,\dots,M-1\}, \\
\Prob\{c_k=M\}=\Prob\{M\in\mathcal{C}_k\}=1,
\end{gather*}
 and the controller is collocated with the actuator (with probability
one).  This essentially corresponds to the conclusions made
by the previous works \cite{gooque08,robkum08} for alternative  NCS configurations without feedback
of plant inputs.\hfs   
\end{ex}

\begin{ex}
\label{ex:M32}
  Consider  $M=3$  and suppose that
  Assumption~\ref{ass:processes} holds. In this case, Proposition~\ref{lem:distr}
  establishes that 
  \begin{equation*}
    \Prob\{c_k=i\}=
    \begin{cases}
      q_1p^2 &\text{if $i=1$,}\\
      pq_2(1-pq_1)&\text{if $i=2$,}\\
      1-pq_2-p^2q_1+p^2q_1q_2&\text{if $i=3$}.
    \end{cases}
  \end{equation*} 
  Note that, since $M$ is small, this result  can alternatively be
  obtained 
  by examining the probabilities of all possible transmission outcomes. This is
  illustrated in Table~\ref{tab:M32}. Of course, for a large number of nodes,
  such a procedure is non-practical and use of Proposition~\ref{lem:distr} is preferable. \hfs  
\end{ex}

\def\baselinestretch{1} 
\begin{table}[t]
  \caption{Set $\mathcal{C}_k$ (with   location $c_k$ in bold-face) for
    $M=3$, see Example~\ref{ex:M32}}
\label{tab:M32}
  \begin{center}
    \begin{tabular}{| c | c | c | c || c | c | }
      \hline
     $\delta_{k-1}^{(1)}$ & $\gamma_k^{(1)} $ &     $\delta_{k-1}^{(2)}$ & $\gamma_k^{(2)}$&
    $\mathcal{C}_k$, $c_k $ & $\Prob$ \\ \hline\hline
     1 & 1 & any & 1 & $\{\mathbf{1}\}$  & $q_1p^2$  \\  \hline
     1 & 1 & any & 0 & $\{1,\mathbf{3}\}$  & $q_1 p (1-p)$  \\  \hline
     1 & 0 & 1 & 1 & $\{1,\mathbf{2}\}$  & $ q_1 (1-p) q_2 p$  \\  \hline
     1 & 0 & 1 & 0 & $\{1,2,\mathbf{3}\}$  & $ q_1 (1-p) q_2 (1-p)$  \\  \hline
     1 & 0 & 0 & any & $\{1,\mathbf{3}\}$   & $ q_1 (1-p) (1-q_2)$ \\  \hline
     0 & any & 1 & 1 & $\{\mathbf{2}\}$  & $(1-q_1)q_2 p$   \\  \hline
     0 & any & 1 & 0 &  $\{2,\mathbf{3}\}$  & $(1-q_1)q_2 (1-p)$ \\  \hline
     0 & any & 0 & any &  $\{\mathbf{3}\}$ & $(1-q_1)(1-q_2)$ \\  \hline
    \end{tabular}
  \end{center}
\end{table}
\def\baselinestretch{1} 

\section{Closed Loop Model}
\label{sec:further-analysis}
The algorithm proposed  embodies a  network driven  distributed state estimation and
control architecture. Closed loop dynamics
 depend upon transmission outcomes,
the plant model~(\ref{eq:1}) and 
nominal controller/estimator dynamics, see~(\ref{eq:11})--(\ref{eq:29}).  
To derive a compact model, it is convenient to introduce the aggregated state estimation vector
$   \hat{x}_k\eq \col\big( \hat{x}_{k}^{(1)}, \hat{x}_{k}^{(2)},\dots  \hat{x}_{k}^{(M)}\big)\in \R^{Mn}.
$ We also denote the ``backup value'' for $u_k$ used at node $i$ as
\begin{equation*}
  \nu_k^{(i)}=
  \begin{cases}
     L\hat{x}_k^{(i)} &\text{if $\mu_k^{(i)}=0$,}\\
    L\hat{x}_k^{(j)}, \,j=c_k^{(i)} &\text{if $\mu_k^{(i)}=1$,}
  \end{cases}
\end{equation*}
see~\eqref{eq:9} and note that $\nu_{k}^{(1)}=L\hat{x}_{k}^{(1)}$, for all $k\in
\N_0$. In view of~(\ref{eq:36}), we have
\begin{equation}
  \label{eq:37}
    \nu_k^{(i)}
    = b_k^{(i)}\hat{x}_k,
\end{equation}
where $b_k^{(1)}\eq e_1\otimes L\in \R^{m\times Mn}$, whereas for $i\geq 2$,
\begin{equation}
\label{eq:43}
  \begin{split}
    b_k^{(i)}&\eq e_\ell \otimes L \in \R^{m\times Mn}\\
    \ell &=
    \begin{cases}
      i &\text{if $c_k^{(i-1)}=0 \lor \gamma_k^{(i-1)}=0$,}\\
      c_k^{(i-1)}&\text{if $c_k^{(i-1)}>0 \land \gamma_k^{(i-1)}=1$,}
    \end{cases} 
  \end{split}
\end{equation}
 depends on the realization of  $\beta_{k-1}$, see~(\ref{eq:32}).

\par
 Since the algorithm gives $u_k=\nu_k^{(M)}$, the plant input estimates used by
 the state estimators  satisfy: 
\begin{equation}
  \label{eq:22}
   \begin{split}
     \hat{u}_{k}^{(i)}&=
    \begin{cases}
    \nu_k^{(M)} &\text{if $\delta_{k}^{(i)}=1$,}\\
      \nu_{k}^{(i)}&\text{if $\delta_{k}^{(i)}=0$}
    \end{cases}\\
    &= \big( \delta_k^{(i)} (e_M\otimes I_m) + \big(1-\delta_{k}^{(i)} \big)
    (e_i\otimes I_m) \big)\nu_{k}, 
  \end{split}
\end{equation}
where $ \nu_k\eq
  \col\big( \nu_{k}^{(1)}, \nu_{k}^{(2)}, 
\dots ,  \nu_{k}^{(M)}\big)\in\R^{Mm}$
forms part of the internal variables used by the $M$ state estimators.

\par Now, the plant model
can be  written as
\begin{equation}
  \label{eq:26}
  x_{k+1}=Ax_k+B\nu_k^{(M)}+w_k
\end{equation}
and~(\ref{eq:26}),~(\ref{eq:15}) and~(\ref{eq:22})  then provide: 
\begin{equation}
  \label{eq:20}
  \begin{split}
    \hat{x}_{k+1}^{(i)}&
    = \big( I_n- K_{k+1}^{(i)}  C\big)
      \big(A\hat{x}_{k}^{(i)} + B\hat{u}_{k}^{(i)}\big)\\
      &\;\;+K_{k+1}^{(i)}  \big(CAx_k+CB\nu_k^{(M)}+Cw_k+v_{k+1}\big)\\
 &= K_{k+1}^{(i)} CAx_k+\big( I_n- K_{k+1}^{(i)}  C\big)
      (e_i\otimes A)\hat{x}_{k}\\&\quad +  d_k^{(i)} \nu_k +K_{k+1}^{(i)}
     \big( C w_k + v_{k+1} \big),\\
    d_k^{(i)}&\eq 
        \big(1-\delta_{k}^{(i)}\big)\big( I_n- K_{k+1}^{(i)}
       C\big) (e_i\otimes B) \\
    &\quad+\big(\big(1-\delta_{k}^{(i)}  \big)K_{k+1}^{(i)}
C +  \delta_k^{(i)} I_n
        \big)(e_M\otimes B).
  \end{split}
\end{equation}
If we now introduce 
\begin{equation}
\label{eq:41}
  \Theta_k\eq
  \col\big( 
    x_k,\hat{x}_k,\nu_k\big),\quad
  n_k\eq \col(   w_k, v_{k+1}),
 \end{equation}
and use~(\ref{eq:27}), then~(\ref{eq:20}) becomes
\begin{equation*}
\begin{split}
   \hat{x}_{k+1}^{(i)}&=\mathcal{D}^{(i)}(\beta_k)\Theta_k+\mathcal{E}^{(i)}(\beta_k)
   n_k,\\
    \mathcal{D}^{(i)}(\beta_k)&\eq
    \begin{bmatrix}
      \Gamma_{k+1}^{(i)} K CA\; & \big( I_n- \Gamma_{k+1}^{(i)} K C\big)
      (e_i\otimes A)\; & d_k^{(i)}
    \end{bmatrix}\\
\mathcal{E}^{(i)}(\beta_k)&\eq\Gamma_{k+1}^{(i)} K 
   \begin{bmatrix}
     C\; & I_p 
   \end{bmatrix}.
\end{split}
\end{equation*}
State estimators, thus, follow the dynamic relation
\begin{equation}
  \label{eq:39}
  \hat{x}_{k+1}=\mathcal{D}(\beta_k)\Theta_k+\mathcal{E}(\beta_k)n_k
\end{equation}
\begin{equation*}
  \mathcal{D}(\beta_k)\eq
  \begin{bmatrix}
    \mathcal{D}^{(1)}(\beta_k) \\ \vdots\\ \mathcal{D}^{(M)}(\beta_k)
  \end{bmatrix},\quad
  \mathcal{E}(\beta_k)\eq
\begin{bmatrix}
    \mathcal{E}^{(1)}(\beta_k)\\ \vdots\\ \mathcal{E}^{(M)}(\beta_k)
  \end{bmatrix}.
\end{equation*}
On the other hand,~(\ref{eq:37}) provides  
\begin{equation}
  \label{eq:40}
  \nu_{k+1}= \mathcal{F}(\beta_k) \Theta_k +  \mathcal{G}(\beta_k) n_k,
\end{equation}
\begin{equation*}
  \mathcal{F}(\beta_k) \eq
  \begin{bmatrix}
    b_{k+1}^{(1)} \mathcal{D}(\beta_k)\\
     \vdots\\
     b_{k+1}^{(M)} \mathcal{D}(\beta_k)
  \end{bmatrix},\quad
\mathcal{G}(\beta_k) \eq
  \begin{bmatrix}
    b_{k+1}^{(1)} \mathcal{E}(\beta_k)\\
     \vdots\\
     b_{k+1}^{(M)} \mathcal{E}(\beta_k)
  \end{bmatrix}.
\end{equation*}
Expressions~(\ref{eq:26}),~(\ref{eq:39}) and~(\ref{eq:40}) lead to the
jump-linear model  
\begin{equation}
  \label{eq:42}
  \Theta_{k+1} = \mathcal{A}(\beta_k) \Theta_k + \mathcal{B}(\beta_k) n_k,
\end{equation}
\begin{equation*}
  \mathcal{A}(\beta_k)\eq
  \begin{bmatrix}
    \begin{bmatrix}
      A\; & 0_{Mn}\; & e_M\otimes B
    \end{bmatrix}\\
      \mathcal{D}(\beta_k)\\
      \mathcal{F}(\beta_k)
  \end{bmatrix},\;
\mathcal{B}(\beta_k)\eq
\begin{bmatrix}
  \begin{bmatrix} I_n\; & 0
  \end{bmatrix}
  \\ \mathcal{E}(\beta_k)\\ \mathcal{G}(\beta_k)
\end{bmatrix}.
\end{equation*}
\begin{ex}
  Consider   $M=2$, in which case
  $\beta_k = 2\gamma_{k+1}^{(1)}+  \delta_{k}^{(1)} $ and $\I=\{0,1,2,3\}$, see~(\ref{eq:32}). Since $c_k^{(1)}
  =\delta_{k-1}^{(1)}$, $\delta_{k-1}^{(2)}=1$ and $b_{k+1}^{(1)} = e_1\otimes L$ for all
  $k\in\N_0$, ~(\ref{eq:43}) yields:
  \begin{equation*}
     b_{k+1}^{(2)}=  
    \begin{cases}
      e_2 \otimes L &\text{if $\beta_{k}<3$,}\\
      e_1 \otimes L &\text{if $\beta_{k}=3$.}
    \end{cases}
  \end{equation*}
In this case, the matrices in~(\ref{eq:39})   are given by
\begin{equation*}
  \begin{split}
    \mathcal{D}(\beta_k)&=
    \begin{bmatrix}
      K CA\; & \big( I_n- K C\big)
      (e_1\otimes A)\; & d_k^{(1)}\\
      \gamma_{k+1}^{(1)} K CA\; & \big( I_n- \gamma_{k+1}^{(1)} K C\big)
      (e_2\otimes A)\; & e_2\otimes B
    \end{bmatrix}\\
    \mathcal{E}(\beta_k)&=
    \begin{bmatrix}
     KC & K
   \\
\gamma_{k+1}^{(1)}   KC & \gamma_{k+1}^{(1)}  K 
    \end{bmatrix},\\
    d_k^{(1)}&=
    \begin{cases}
      \begin{bmatrix}
         \big( I_n- K 
       C\big)B\; & KCB
      \end{bmatrix}
     &\text{if $\beta_{k} \in\{0,2\}$,}\\
     e_2\otimes B&\text{if $\beta_{k} \in\{1,3\}$}
    \end{cases}
  \end{split}
\end{equation*}
thereby, characterizing the model~(\ref{eq:42}).\hfs
\end{ex}
\section{Performance Analysis}
\label{sec:wnat-next}
To analyze the NCS via~(\ref{eq:42}), we will  adopt
the stochastic modeling framework of \cite{queahl13a}. 
 Transmission outcome 
distributions depend upon the fading radio environment. To allow for temporal
and  spatial correlations of the radio environment (and possibly also for power and
bit-rate control), in \cite{queahl13a} we used a Markovian
\emph{network 
  state}, $\{\Xi_k\}$, $k\in\N_0$, which takes values in a finite set, say
$\B$. Each element of $\B$ corresponds to a possible configuration of the
physical environment, e.g., position of mobile objects. 
Dropout probabilities of individual channels, \emph{when conditioned on the
network state}, are considered independent. In the particular instance
where $\B$ has only one element, the model describes a situation with
 independent Bernoulli channels. For the present purposes, the model can
be summarized via:

\begin{ass}
\label{ass:markov}
The  process $\{\Xi_k\}$, $k\in\N_0$ is an aperiodic homogeneous
  Markov Chain with transition
  probabilities 
  $    p_{ij}=\Prob\{\Xi_{k+1}=j\,|\, \Xi_k=i\}$, $i,j \in\B$ 
and stationary distribution $\pi_i=\lim_{k\to\infty}\Prob\{\Xi_k=i \}$, $i \in\B.$
The aggregated transmission outcome process $\{\beta_k\}$ in~(\ref{eq:32}) is conditionally independent
  given the network state $\{\Xi_k\}$, i.e.,  
  $    \phi_{ij}\eq \Prob\{\beta_k=i\,|\,\Xi_k=j\}$, 
  for all $(i,j)\in\I\times\B$. \hfs
\end{ass}
It is worth noting that, with the above model, the process $\beta_k$ is
correlated, but not necessarily Markovian. However, the
augmented jump
process $(\beta_k,\Xi_k)$, $k\in\N_0$ forms a finite Markov Chain. Thus, under
Assumption~\ref{ass:markov},~(\ref{eq:42}) belongs
to the class 
of Markov jump-linear systems, as studied for example in \cite{cosfra05,leedul07}. In
particular, Theorems 3.9 and 3.33 of \cite{cosfra05} establish necessary and
sufficient 
conditions for mean-square stability (MSS) which can be stated in terms of
feasibility 
of a
linear-matrix inequality.
The
following result characterizes closed loop performance of the flexible networked
control systems architecture of interest in the present work. It is tailored directly
to the model in 
Assumption~\ref{ass:markov} without needing to resort to   the augmented jump
process $(\beta_k,\Xi_k)$.   
\begin{thm}
\label{thm:performance-analysis}
  Suppose that Assumption~\ref{ass:markov} holds, that the system~(\ref{eq:42})
  is MSS and define $W\eq \mathrm{diag}(Q,R)$,
\begin{equation}
  \label{eq:49}
  \begin{split}
    \overline{\mathcal{A}}_j&\eq \E\big\{\mathcal{A}(\beta_k)\,\big
    |\,\Xi_k=j\big\}=
    \sum_{i\in\I}\phi_{ij}\mathcal{A}(i),\quad j\in\B,\\
    \overline{\mathcal{B}}_j&\eq \E\big\{\mathcal{B}(\beta_k)\,\big
    |\,\Xi_k=j\big\}=
    \sum_{i\in\I}\phi_{ij}\mathcal{B}(i),\quad j\in\B.
  \end{split}
\end{equation}
Then
\begin{equation}
  \label{eq:53}
  \lim_{k\to\infty} \E\{\Theta_{k}\Theta_{k}^T\} = \sum_{i\in\B} H_i,
\end{equation}
where $H_i$, $i\in\B$ satisfy the linear system of equations:
\begin{equation}
  \label{eq:52}
  H_{i}=\sum_{j\in\B}p_{ji}\overline{\mathcal{A}}_i
    H_{j} 
    (\overline{\mathcal{A}}_i)^T    + \pi_i\overline{\mathcal{B}}_i
    W(\overline{\mathcal{B}}_i)^T.
\end{equation}
\end{thm}
In view of~(\ref{eq:41}) and the fact that $u_k=\nu_k^{(M)}$, our result can 
be  used to evaluate the plant state and input covariances. 

\begin{rem}
  By using \cite[Sec.5]{lancas70},   $H_i$ in~(\ref{eq:52}) can be written in terms of the solution to a
  system of linear 
  equations. 
\hfs
\end{rem}


\begin{figure}[t!]
\centering
\includegraphics[width=0.4\textwidth]{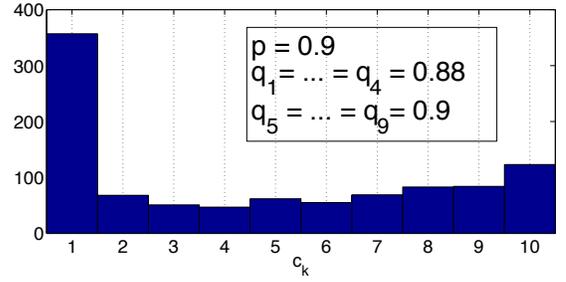}
\caption{Histogram of   $c_k$ for an i.i.d.\ network
  with success probabilities $p=0.9$, $q_1= \dots = q_4=0.88$, and $q_5= \dots=q_9=0.9$.} 
\label{fig:location3}
\end{figure}

\section{Simulation Studies}
\label{sec:simulation-study}
We consider an NCS  with $M=10$ nodes.   In addition to   the implementation of the controller via
Algorithm~\ref{alg1}, we also examine   two baseline NCS
architectures. In the first one, the controller  and estimator are fixed  at the
actuator node: 
  \begin{equation}
    \label{eq:35}
    \begin{split}
      x_{k+1}&=Ax_k +BL\hat{x}^{a}_k +w_k,\\
    \hat{x}^{a}_{k} &= A\hat{x}^{a}_{k-1}  +
    Bu_{k-1}\\
    &\quad+\Gamma_k^{(10)}K\big(y_k - C ( A\hat{x}^{a}_{k-1}  +
    Bu_{k-1})\big),
  \end{split}
\end{equation}
where $\Gamma_k^{(10)}$ is as in~(\ref{eq:27}).
In the second scheme, controller and estimator are implemented at the sensor
node. If the controller output
 is lost, then the previous plant input is held:
\begin{align}
    x_{k+1}&=Ax_k
    +\Gamma_k^{(10)}BL\hat{x}^{s}_k+(1-\Gamma_k^{(10)})Bu_{k-1}+w_k,\notag \\
    \hat{x}^{s}_{k} &= A\hat{x}^{s}_{k-1}  +
    B\hat{u}^{s}_{k-1}  \label{eq:38}\\
    &\quad+K\big(y_k - C ( A\hat{x}^{s}_{k-1} +
    B\hat{u}^{s}_{k-1} )\big),\notag
  \end{align}
  where $\Gamma_k^{(10)}$ is as in~\eqref{eq:27} and
\begin{equation*}
  \hat{u}^{s}_{k-1} =
  \begin{cases}
    u_{k-1} &\text{if $\delta_{k-1}^{(1)}=1$,}\\
     L\hat{x}^{s}_{k-1} &\text{if $\delta_{k-1}^{(1)}=0$.}
  \end{cases}
\end{equation*}


\paragraph*{Independent and identically distributed dropouts}
We first consider i.i.d.\ transmission processes  as
per Assumption~\ref{ass:processes}.
Fig.~\ref{fig:location3}
illustrates a histogram of $c_k$, obtained by running the algorithm
for 1000 steps with dropout probabilities as indicated. Whilst~(\ref{eq:19})
shows  that for small  $p$, control 
calculations are at most times, carried out at the actuator node,
Fig.~\ref{fig:location3} illustrates that  if
links are more reliable, then the controller will be placed at the sensor node at
most time steps. 


 \begin{figure}[t!]
 \centering
\includegraphics[width=0.48\textwidth]{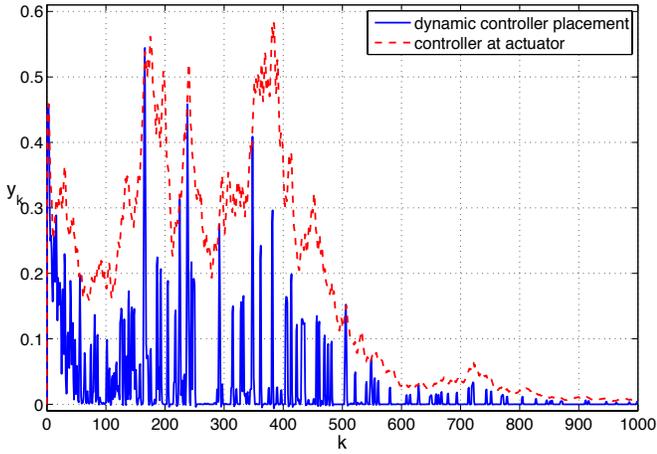}
 \caption{Output trajectory of the plant model~\eqref{unstable} for an i.i.d.\ network
   with success probabilities $p=0.9$, $q_1= \dots = q_9=1$.} 
 \label{fig:salidaUnstable}
 \end{figure}

\par We first consider a noiseless
unstable plant model~(\ref{eq:1}), where 
 \begin{equation}
\label{unstable}
   A=
   \begin{bmatrix}
\begin{array}{cc}
	1.87&-0.86\\
	1&0
\end{array}\
   \end{bmatrix},\; 
  B=
   \begin{bmatrix}
\begin{array}{c}
	1\\
	0
\end{array}
   \end{bmatrix},\; 
  C=
   \begin{bmatrix}
   \begin{array}{cc}
	0.048&0.045
\end{array}
   \end{bmatrix}
 \end{equation}
 with Gaussian initial state having mean $\bar{x}_0 =
 [\begin{matrix}
   5 & 5
 \end{matrix}]^T$ and covariance $P_0=0.1\times I_2$. Controller and
 estimator  gains 
 $L$ and $K$ correspond to the steady state  LQG/LQR controller with stage cost
 $\|x_k\|^2 + \|u_k\|^2/10$. All nodes use
 as initial state estimates, $\hat{x}^{(i)}_0=[\begin{matrix}
   0 & 0
 \end{matrix}]^T$. The network has i.i.d.\ dropouts   with success probabilities $p=0.9$ and    $q_1=
 \dots  =q_9=1$. 
 \par  The baseline NCS~\eqref{eq:38} failed to stabilize the
present system.
Fig.~\ref{fig:salidaUnstable} compares a typical  plant output trajectory obtained by using the
 proposed algorithm (solid line) with that provided by the baseline NCS~\eqref{eq:35} (dashed line).
   As can be appreciated, the adaptive
controller allocation algorithm presented reacts
more quickly to plant outputs. It thereby recovers more quickly from the very
bad local 
initial state estimates and provides control actions leading to
 faster convergence to the origin.
If we adopt the empirical performance measure
 \begin{equation}
\label{eq:24}
   J\eq  \sum_{k=1}^{1000}y_k^2, 
 \end{equation}
then, with the dynamic architecture, we obtain $J\approx 3$, whereas for the baseline
NCS described by~\eqref{eq:35}, $J\approx 11$.

\par We next consider a plant model with an integrator, where
 \begin{equation}
\label{eq:2}
   A=
   \begin{bmatrix}
\begin{array}{cc}
	1.8&-0.8\\
	1&0
\end{array}\
   \end{bmatrix}\!,\; Q=0.01\times I_2,\; R=0.01
 \end{equation}
and $B$ and $C$ as
in~(\ref{unstable}). The 
 initial state has mean $\bar{x}_0 =
 [\begin{matrix}
   10 & 10
 \end{matrix}]^T$. Table \ref{table2} illustrates how the 
performance gained by using the proposed method depends upon the network reliability. For the situation
examined, larger performance gains are obtained with smaller $p$. For larger $p$,
the performance gains become less relevant. This  finding
is intuitive, since for $p\approx 1$ the network becomes transparent and overall
performance is dominated by the nominal design~(\ref{eq:1})--(\ref{eq:29}).

\begin{table}[t]
\caption{Performance indices $J$ when controlling the system~\eqref{eq:2} over an i.i.d.\ network with 
  $q_1= \dots= q_4=0.99$, $q_5= \dots=q_9=0.995$.} 
\label{table2} 
\begin{center}\begin{tabular}{c|c|c|c}
$p$ & NCS~(\ref{eq:12})--(\ref{eq:15}) & NCS~\eqref{eq:35} & NCS~\eqref{eq:38}\\
\hline 
0.95 & 33.5 & 38.5   &  (unstable)\\
\hline 
0.97 &   33.2 & 37.6   &  56.1  \\
\hline 
0.99 &  33.1 & 34.7   & 34.1 
\end{tabular}
\end{center}
\end{table}

\begin{figure}[t]
 \centering
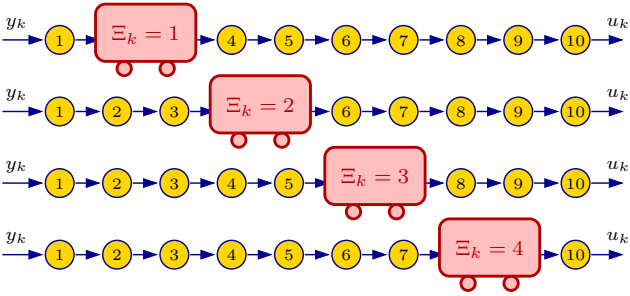
 \caption{Sensor-actuator network with  moving obstacle.} 
 \label{fig:rednoiid}
 \end{figure}

\begin{figure}[t]
 \centering
\includegraphics[width=0.45\textwidth]{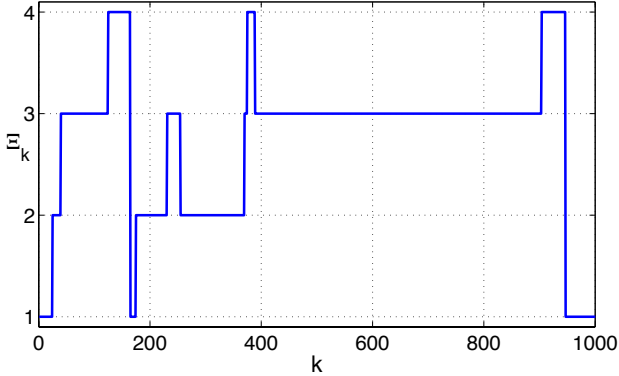}
 \caption{Network state trajectory, $\Xi_k$.} 
 \label{fig:xik}
 \end{figure}

\begin{figure}[t]
 \centering
\includegraphics[width=0.45\textwidth]{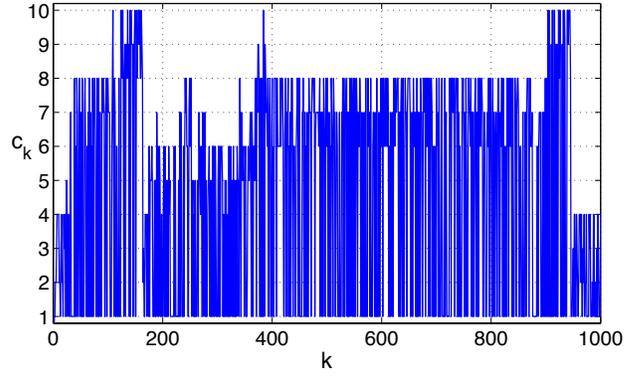}
 \caption{Controller location $c_k$ for the network in Fig.~\ref{fig:rednoiid}.} 
 \label{fig:CkNoIID}
 \end{figure}

\paragraph*{Network with moving obstacle}
We  now focus on a   network with an obstacle (e.g.,
a 
robot or crane) moving between four 
different positions, see Fig.~\ref{fig:rednoiid}. We model this  as
in Section~\ref{sec:wnat-next}, using the network state process
$\Xi_k\in\B=\{1,2,3,4\}$. The transition probabilities for $\Xi_k$ are given by:
\begin{equation*}
  [p_{ij}]=
  \begin{bmatrix}
      0.99 &  0.01 & 0 & 0\\ 
   0.003 &  0.99 & 0.007 & 0\\ 
   0 &  0.003 & 0.99 & 0.007\\ 
   0.007 &  0 & 0.003 & 0.99\\ 
  \end{bmatrix}.
\end{equation*}
 The individual link reliabilities
depend on the position of the obstacle. Nodes which are not
blocked benefit from high success probabilities   $r\in[0.88,1]$. Due to
the obstacle, some of the 
success probabilities will, at times, be lowered to $60\%$:

\begin{equation}
\label{eq:25}
\begin{split}
   \Prob\{\gamma_k^{(i)}=1\,|\, \Xi_k=j\} &=
    \begin{cases}
      0.6 &\text{if $i \in\{2j-1,2j,2j+1\}$}\\
      r&\text{in all other cases}
    \end{cases}\\
    \Prob\{\delta_{k-1}^{(i)}=1\,|\, \Xi_k=j\} &=\begin{cases}
      0.6 &\text{if $i \in\{2j,2j+1\}$}\\
      r&\text{in all other cases.} 
    \end{cases}
  \end{split}
\end{equation}
For $r=0.99$,  Figs.~\ref{fig:xik} and~\ref{fig:CkNoIID} illustrate how using
 Algorithm~\ref{alg1} the controller
location depends upon the network state $\Xi_k$. It turns out that, in the
present case, the plant input is 
 always provided by one of the nodes located between the sensor node and the
 node immediately following the blocked ones. This behaviour can be explained by noting
 that, in absence of the obstacle, the network is very reliable. In fact, if
 none of the nodes were 
blocked, then the algorithm would (almost) always locate the controller
 at the sensor node. 


\begin{figure}[t]
 \centering
\includegraphics[width=0.44\textwidth]{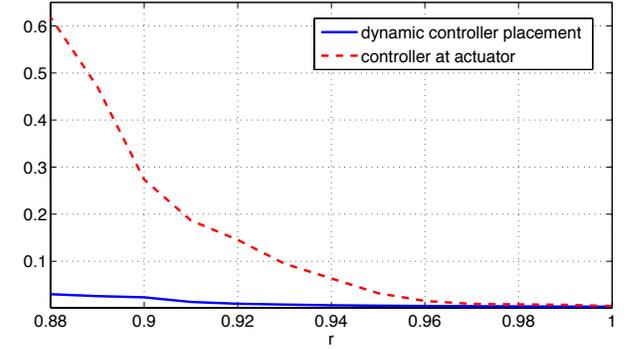}
 \caption{Trace of the stationary covariance of the plant state $x_k$
   in~(\ref{eq:2}) as a
   function of the success probability $r$, see~(\ref{eq:25}).} 
 \label{fig:cov}
 \end{figure}

For~\eqref{eq:2} and with $r=0.99$, use of the dynamic
architecture proposed, gave $J=51.6$. In contrast, if the controller is fixed at the
actuator node, see~(\ref{eq:35}), then $J= 68.5 $ was obtained. This amounts to a
performance loss of $33\%$. In the situations examined, 
positioning  the baseline NCS~\eqref{eq:38} failed to stabilize the plant model.
Fig.~\ref{fig:cov}, obtained  using Theorem~\ref{thm:performance-analysis}, illustrates 
how the trace of the covariance of the plant state~\eqref{eq:2} depends on the network
parameter $r$ in~(\ref{eq:25}). In this figure, the solid line corresponds to the
  dynamic controller placement method, whereas the dashed line refers to
the baseline controller~(\ref{eq:35}). Our results  clearly indicate that,
without the need for controller re-design, the algorithm proposed
in the present work has the potential to give  significant performance gains
when compared to earlier NCS configurations where node functionalities are fixed.

\section{Conclusions}
\label{sec:conclusions}
We have presented a flexible architecture for the implementation of a linear
control law  over a wireless sensor-actuator network using analog 
erasure channels without acknowledgments. With the algorithm provided,   the
role played by individual 
nodes  depends on transmission outcomes. In particular, the controller
location is adapted to the availability of past plant input values and
transmission outcomes. By deriving a Markovian jump-linear system model, we
established a closed form expression for the stationary covariance of the system
state in the presence of 
correlated dropout processes. Future work may include extending the ideas  to   multiple-loops, to general network topologies,  and to controller
design.  

\bibliography{/Users/daniel/Dropbox/dquevedo}

\appendix

\section{Proof of Proposition~\ref{lem:distr}}
\label{sec:proof-lemma}
With
Assumption~\ref{ass:processes},   $\Prob\{ \mu_k^{(1)}=1\} =\Prob\{
\delta_{k-1}^{(1)}=1\} =q_1.$ 
It is easy to see from lines~\ref{alg:empty} and~\ref{alg:fwd} of
Algorithm~\ref{alg1} that 
\begin{equation}
  \label{eq:10}
  \begin{split}
    i&\in\mathcal{C}_k \Longleftrightarrow \delta_{k-1}^{(i)} = 1 \\
    &\quad\land \big( 
    \gamma_k^{(i-1)}=0 \lor \big(  \gamma_k^{(i-1)}=1\land \mu_k^{(i-1)}=0
    \big)\big)\\ 
    &\Longleftrightarrow \delta_{k-1}^{(i)} = 1\land \big( 
    \gamma_k^{(i-1)}=0 \lor \mu_k^{(i-1)}=0
    \big)
  \end{split}
\end{equation}
and that, similarly
\begin{equation}
\label{eq:14}
\begin{split}
    \mu_k^{(i)}&=0 \Longleftrightarrow \big(\gamma_k^{(i-1)}=0\land \delta_{k-1}^{(i)} =0
     \big)\\
    & \lor \big( \gamma_k^{(i-1)}=1\land
    \mu_k^{(i-1)}=0 \land  \delta_{k-1}^{(i)} =0 \big)\\
     &\Longleftrightarrow  \delta_{k-1}^{(i)} =0 \\
     &\land \big( 
    \gamma_k^{(i-1)}=0\lor \big(\gamma_k^{(i-1)}=1\land\mu_k^{(i-1)}=0  \big)\big)\\
     &\Longleftrightarrow  \delta_{k-1}^{(i)} =0 \land \big( 
    \gamma_k^{(i-1)}=0\lor \mu_k^{(i-1)} =0  \big)
  \end{split}
\end{equation}
for all $i\in \{1,2,\dots,M\}$. 
(\ref{eq:14}) provides the recursion
\begin{equation*}
  \begin{split}
  \Prob&\big\{ \mu_k^{(i)}=1\big\} =1-\Prob\big\{\delta_{k-1}^{(i)} =0 \big\}\\
  &\qquad \times\Prob\big\{\mu_k^{(i-1)}=0 \lor \gamma_k^{(i-1)}=0 \big\}\\
  &=1- (1-q_i) \big( 1-  \Prob\big\{\mu_k^{(i-1)}=1 \land \gamma_k^{(i-1)}=1 \big\}\big)\\
  &=1-(1-q_i) \big( 1-  p\Prob\big\{\mu_k^{(i-1)}=1\big\}  \big)\\
  &=q_i+p(1-q_i)\Prob\big\{\mu_k^{(i-1)}=1\big\},
\end{split}
\end{equation*}
having explicit solution~(\ref{eq:18}).
On the other hand,~(\ref{eq:10}) gives
\begin{equation*}
  \begin{split}
  &\Prob\{  i\in\mathcal{C}_k \} 
  \\
  &= \Prob\big\{ \delta_{k-1}^{(i)} = 1 \big\} \big(
    1 -\Prob\big\{ \gamma_{k}^{(i-1)} = 1 \big\}\Prob\big\{ \mu_{k}^{(i-1)} = 1
    \big\} \big),
  \end{split}
\end{equation*}
thus establishing~(\ref{eq:16}).
 By~(\ref{eq:4}) we obtain
\begin{equation*}
  \begin{split}
    \Prob\{c_k=i\} &= \Prob\{\max (\mathcal{C}_k)=i\}\\
    &= \Prob \{i\in\mathcal{C}_k
    \land \gamma_k^{i}=\gamma_k^{i+1}=\dots =\gamma^{M-1}=1 \}\\
  \end{split}
\end{equation*}
for $i\in\{1,\dots,M-1\}$, whereas for the actuator node, we have
$\Prob\{c_k=M\} = \Prob
\{M\in\mathcal{C}_k\}$. This proves~(\ref{eq:21}). \hfs

\section{Proof of Theorem~\ref{thm:performance-analysis}}
\label{app:proof-theor-refthm:p}
By the law of total expectation, we have
 \begin{equation}
   \label{eq:31}
   \E\{\Theta_{k+1}\Theta_{k+1}^T\} = \sum_{i\in\B} H_{k+1,i}, 
 \end{equation}
where
\begin{equation}
  \label{eq:33}
   H_{k+1,i} \eq \E\{\Theta_{k+1}\Theta_{k+1}^T\,|\,\Xi_k=i\} \Prob\{\Xi_k=i\}.
\end{equation}
Now, the system equation~(\ref{eq:42}) together with the network fading model in
Assumption~\ref{ass:markov} allow one to write
\begin{align}
    \E&\{\Theta_{k+1}\Theta_{k+1}^T\,|\,\Xi_k=i\} =
    \E\big\{\big(\mathcal{A}(\beta_k) \Theta_k + \mathcal{B}(\beta_k) n_k\big)
    \notag \\
    &\quad\times\big(\mathcal{A}(\beta_k) \Theta_k + \mathcal{B}(\beta_k)
    n_k\big)^T\,\big|\,\Xi_k=i\big\} \notag \\
    &= \E\big\{ \mathcal{A}(\beta_k)
    \Theta_k\Theta_k^T\mathcal{A}(\beta_k)^T\,\big|\,\Xi_k=i\big\} \label{eq:34}
    \\
    &\quad+\E\big\{ \mathcal{B}(\beta_k)
    n_kn_k^T\mathcal{B}(\beta_k)^T\,\big|\,\Xi_k=i\big\} \notag\\
    &= \E\big\{ \mathcal{A}(\beta_k)
    \Theta_k\Theta_k^T\mathcal{A}(\beta_k)^T\,\big|\,\Xi_k=i\big\} + \overline{\mathcal{B}}_i
    W(\overline{\mathcal{B}}_i)^T\notag
  \end{align}
  since $\{n_k\}$ is zero-mean i.i.d. The rule of total expectation,  Bayes'
  rule and the Markovian property of~(\ref{eq:42}) give that  
\begin{align}
\E&\big\{ \mathcal{A}(\beta_k)
    \Theta_k\Theta_k^T\mathcal{A}(\beta_k)^T\,\big|\,\Xi_k=i\big\}\notag\\
    &= \sum_{j\in\B}\E\big\{ \mathcal{A}(\beta_k)
    \Theta_k\Theta_k^T\mathcal{A}(\beta_k)^T\,\big|\,\Xi_k=i,\Xi_{k-1}=j\big\}\notag\\
    &\quad\times
    \Prob \{\Xi_{k-1}=j\,|\,\Xi_k=i \}\label{eq:47}\\
    &= \sum_{j\in\B} \E\big\{ \mathcal{A}(\beta_k)
    \Theta_k\Theta_k^T\mathcal{A}(\beta_k)^T\,\big|\,\Xi_k=i,\Xi_{k-1}=j\big\}\notag\\
    &\quad\times
    \Prob \{ \Xi_{k}=i\,|\,\Xi_{k-1}=j \}  \Prob\{\Xi_{k-1}=j \}/\Prob\{\Xi_k=i
    \}\notag\\
    &=\!\sum_{j\in\B} p_{ji}\overline{\mathcal{A}}_i
    \E\big\{\Theta_k\Theta_k^T\,\big |\, \Xi_{k-1}=j\big\} 
    (\overline{\mathcal{A}}_i)^T       \frac{ \Prob\{\Xi_{k-1}=j
      \}}{\Prob\{\Xi_k=i    \}}. \notag
  \end{align}
  Substitution of~(\ref{eq:47}) into~(\ref{eq:34}) and then into~(\ref{eq:33})
  provides 
\begin{align}
    H_{k+1,i} &=\sum_{j\in\B}p_{ji}\overline{\mathcal{A}}_i
    \E\big\{\Theta_k\Theta_k^T\,\big |\, \Xi_{k-1}=j\big\} 
    (\overline{\mathcal{A}}_i)^T \label{eq:50}\\ 
    &\qquad \times \Prob\{\Xi_{k-1}=j \}+ \overline{\mathcal{B}}_i
    W(\overline{\mathcal{B}}_i)^T\Prob\{\Xi_k=i \}\notag \\
    &=\sum_{j\in\B}p_{ji}\overline{\mathcal{A}}_i
    H_{k,j} 
    (\overline{\mathcal{A}}_i)^T    + \overline{\mathcal{B}}_i
    W(\overline{\mathcal{B}}_i)^T\Prob\{\Xi_k=i \}\notag
  \end{align}
  Since   the NCS is assumed MSS, it is  asymptotically wide-sense
stationary \cite[Thm.\ 3.33]{cosfra05}. If we  define
$ H_i\eq\lim_{k\to \infty} H_{k,i}$, $i\in\B$,
and recall that $\{\Xi_k\}$   is aperiodic, then~(\ref{eq:50})
becomes~(\ref{eq:52}), and~(\ref{eq:31}) 
establishes~(\ref{eq:53}).\hfs

\end{document}